\documentclass[twocolumn,superscriptaddress,amsmath,amssymb]{revtex4}
\usepackage[dvips]{graphicx}
\usepackage{dcolumn}
\usepackage{bm}

\def\bd{
\begin{document}} \def\ed{\end{document}}
\def\bmp{\begin{minipage}} \def\emp{\end{minipage}}
\def\bcc{\begin{center}} \def\ecc{\end{center}}     \def\npg{\newpage}
\def\beq{\begin{equation}} \def\eeq{\end{equation}} \def\hph{\hphantom}
\def\be{\begin{equation}} \def\ee{\end{equation}} \def\r#1{$^{[#1]}$}
\def\n{\noindent} \def\ni{\noindent} \def\pa{\parindent}
\def\hs{\hskip} \def\vs{\vskip} \def\hf{\hfill} \def\ej{\vfill\eject}
\def\cl{\centerline} \def\ob{\obeylines}  \def\ls{\leftskip}
\def\underbar#1{$\setbox0=\hbox{#1} \dp0=1.5pt \mathsurround=0pt
   \underline{\box0}$}   \def\ub{\underbar}    \def\ul{\underline}
\def\f{\left} \def\g{\right} \def\e{{\rm e}} \def\o{\over} \def\d{{\rm d}}
\def\vf{\varphi} \def\pl{\partial} \def\cov{{\rm cov}} \def\ch{{\rm ch}}
\def\la{\langle} \def\ra{\rangle} \def\EE{e$^+$e$^-$} \def\pt{p_{\rm t}}
\def\bitz{\begin{itemize}} \def\eitz{\end{itemize}}
\def\btbl{\begin{tabular}} \def\etbl{\end{tabular}}
\def\btbb{\begin{tabbing}} \def\etbb{\end{tabbing}}
\def\beqar{\begin{eqnarray}} \def\eeqar{\end{eqnarray}}
\def\\{\hfill\break} \def\dit{\item{-}} \def\i{\item}
\def\bbb{\begin{thebibliography}{9} \itemsep=-1mm}
\def\ebb{\end{thebibliography}} \def\bb{\bibitem}
\def\bpic{\begin{picture}(260,240)} \def\epic{\end{picture}}
\def\akgt{\cl{\bf ACKNOWLEDGMENTS}}
\def\fgn{\noindent{\bf\large\bf figure captions}}
\def\lan{\langle}
\def\ran{\rangle}
\def\p{\pi}
\def\ifmath#1{\relax\ifmmode #1\else $#1$\fi}%
\def\rc{\ifmath{{\mathrm{c}}}}
\def\cut{\ifmath{{\mathrm{cut}}}}
\def\rF{\ifmath{{\mathrm{F}}}}
\def\rK{\ifmath{{\mathrm{K}}}}
\def\rp{\ifmath{{\mathrm{p}}}}
\def\rt{\ifmath{{\mathrm{t}}}}
\def\LAB{\ifmath{{\mathrm{LAB}}}}
\def\cut{\ifmath{{\mathrm{cut}}}}
\def\beq{\begin{equation}}
\def\eeq{\end{equation}}

\newcommand{\cinst}[2]{$^{\mathrm{#1}}$~#2\par}
\newcommand{\crefi}[1]{$^{\mathrm{#1}}$}
\newcommand{\crefii}[2]{$^{\mathrm{#1,#2}}$}
\newcommand{\crefiii}[3]{$^{\mathrm{#1,#2,#3}}$}
\newcommand{\HRule}{\rule{0.5\linewidth}{0.5mm}}

\bd

\title{Initial orientation effect and selecting desired events\\ in 520AMeV/u U-U collisions}

\author{K.J. Wu}
\affiliation{ Institute of Particle Physics, Hua-Zhong Normal
University, Wuhan 430079, China}

\affiliation{ Key Laboratory of
Quark $\&$ Lepton Physics (Huzhong Normal University), Ministy of
Education, China}

\author{F. Liu}
\affiliation{ Institute of Particle Physics, Hua-Zhong Normal
University, Wuhan 430079, China}

\affiliation{ Key Laboratory of Quark $\&$ Lepton Physics (Huzhong
Normal University), Ministy of Education, China}

\author{N. Xu}
\affiliation{ Nuclear Science Division, Lawrence Bekeley National
Laboratory, Bekeley, CA 94720,USA}

\begin{abstract}
How to select out those collisions with the desired geometry such as
tip-tip and/or body-body in experiment is one key point for
performing high energy UU collisions. With a relativistic transport
model, we performed a simulation for deformed UU collision with vast
different orientations at CSR energy area corresponding to the high
net-baryon density region in QCD phase diagram. By investigating the
centrality and initial collision orientation dependence of the
center baryon density, we found that the tip-tip like UU collisions
with extended high density phase, which is very important for
studying the nuclear EoS of high baryon density matter and the
possible end-point of the phase boundary, are those events with
small initial orientations ($\leq20^{0}$) for bath projectile and
target in reaction plane and small impact parameter ($\leq2.6fm$).
We pointed out quantificationally two observations -- multiplicity
of forward neutron and nuclear stopping power that both allows us to
select out those most interesting events (i.e. tip-tip like), which
will be very helpful for the future experiments at performing UU
collisions.

\vskip 3mm \normalsize\sl \noindent PACS: \ 21.\,60.\,Ka,
21.\,10.\,Re, 24.\,10.\,Lx
\par
\vskip 3mm

\end{abstract}

\maketitle

\section{\bf Introduction}
A major goal of current and future high energy heavy-ion collisions
experiments is to probe and study the state of new matter under
extreme conditions. Lattice QCD calculation predicts that a phase
transition from the hadronic state to the quark-gluon-plasma (QGP)
will occur under high temperature or density.

In recent years, the focus on searching for QGP is performed at
SPS/CERN and RHIC/BNL with ultra-relativistic heavy ion collision
correspond to high temperature and low baryon density region in
nuclear matter phase diagram~\cite{Q1,Q2}. Several experimental
observables like the Number-of-Constituent-Quark (NCQ) Scaling of
elliptic flow, jet-quenching, etc. can be explained commendably with
the appearance of partonic degree of freedom in the collisions at
RHIC energy regions~\cite{U0}. However, due to the complex nature of
the relativistic nucleus-nucleus reactions, the QGP, if it has been
created, escape direct detection. The fact we have not observed any
observation undergoing dramatic change reminds us to perform an
energy scan from high energy to lower energy to search for the phase
boundary and the possible end-point of the phase boundary~\cite{Q3}.
On the other hand, in last two decades the heavy ion collisions
performed at the BEVALAC/LBNL and SIS/GSI~\cite{U1,U2} were used to
produce hot and compressed nuclear matter for studying more about
the nuclear equation of state (EoS)~\cite{U3,U4} at high baryon
density and low temperature region of the phase diagram. We have
made great efforts in studying the nuclear EoS, both theoretically
and experimentally, but a solid conclusion can hardly be made. From
this two points for more understanding of the nuclear matter phase
diagram and EoS at high net-baryon density region, it is expected
that the Heavy Ion Research Facility in Lanzhou china (HIRFL)---
Cooler Storage Ring (CSR) with a maximal beam kinetic energy of 520
MeV/nucleon for heavy nuclei~\cite{U5}, which focuses at the high
baryon density region, can make significant contribution to those
important search.

The uranium-on-uranium (UU) collision is most suitable for this
study. Uranium is the most deformed stable nucleus. Representing U
as a homogeneous ellipsoid with one long ($R_{l}$) and two short
($R_{s}$) semi-axis, one can related their ratio to deformation
parameter used in nuclear physics
\beqar  
\frac{R_{l}}{R_{s}}= (\frac{1+4\delta/3}{1-2\delta/3})^{1/2} \eeqar
For $\delta_{U}\approx0.27$, the ratio of the long-axis over
short-axis is as large as 1.29. For A=238, we will use $R_{l}=8.4$,
$R_{s}=6.5$ and $R=({R_{s}^{2}R_{l}})^{1/3}=7.0 fm$~\cite{Q4}.

Due to larger A and deformation, the gain in energy density for UU
over AuAu can reach the factor 1.8 in RHIC/BNL~\cite{S1}. Because of
the deformation, UU collisions at the same beam energy and impact
parameter but different orientations are expected to form dense
matter with different compressions and lifetimes.

We consider two extreme collisions: the head on collisions with
long-axis on long-axis and short-axis on short-axis as tip-tip and
body-body collisions, respectively~\cite{U7}. Random collision
geometries, which are illustrated in Fig.1, lie between them. It is
expected that the tip-tip collisions can form a higher densities of
nuclear matter with longer duration than in body-body or the
spherical nuclei collisions and easy to reach thermal equilibrium at
the same energy and impact parameter. This is a powerful tool for
studying the physics of large compression, high-baryon density and
possible phase transition from nuclear matter to a new form of
matter with partonic degree of freedom.

\begin{figure}[tbph]
\resizebox{!}{50mm}{\includegraphics{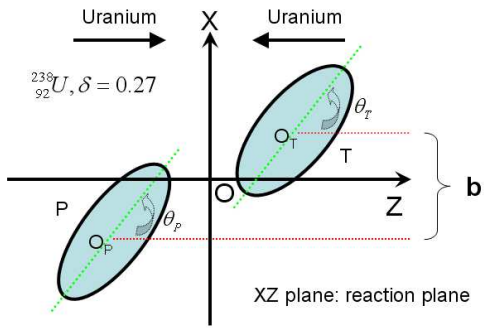}}
\caption{\label{Fig.1} The initial collision geometries of the
reaction plane in U+U collisions: Z-direction is defined as the beam
direction. The impact parameter b is along X axis. X-Z plane
represents the reaction plane. $\theta_{P}$ and $\theta_{T}$ are the
orientation of reaction plane for projectile and target,
respectively. Thus, $\theta_{P}=\theta_{T}=0^{0}/\pm90^{0}$
corresponded to tip-tip/body-body collisions respectively.}
\end{figure}

For the non-polarized UU collision, target and projectile have
random orientation at the initial coordinate space. Several ideas
have been developed during the past few years to take advantage of
the UU collisions~\cite{Q5,Q6,Q7}, but no experimental
implementation has been made. One of the uncertainties is how to
select collision with the desired geometry such as tip-tip and/or
body-body.

In previous Ref~\cite{U7,U8}, we have studied systemically the
stopping power and anisotropy flow in tip-tip and body-body UU
collisions. In this paper, we apply a relativistic transport models,
ART~\cite{Q8}, to study the effect of different colliding
orientations and only focus on the central baryon density in order
to confirm quantificationally that what initial geometries are the
interesting events with longtime and high density. This will provide
a useful help for quantitative analysis in experiment.

The outline of this paper is as follows. Section II is devoted to
the centrality dependence of central baryon density. In section III,
we will discuss the effect with different inial collision
orientations. In section IV, two measurable ways to select out the
tip-tip like events will been discussed. Finally, a short summary
will be given in section V.

\section{\bf The centrality dependence of central baryon density}
To form a quark-gluon plasma, it is necessary to achieve a high
local energy or baryon densities in a sufficiently large volume and
for a sufficiently long time so that the initial  plasma droplets
can grow up. The current estimate for the critical baryon and energy
densities at which the QGP may form are about 5 $\rho_{0}$ and 2.5
GeV/fm$^{3}$, respectively~\cite{Q9}. Under the assumption of full
nuclear stopping, which is a good approximation for heavy nuclei at
CSR energy region, as the results of energy and baryon number
conservation, the maximum energy density and maximum baryon density
for heavy nuclear collisions are
\beqar  
\frac{\epsilon_{max}}{\epsilon_{0}}= 2.5;
\frac{\rho_{max}}{\rho_{0}} = 2.5, \eeqar respectively~\cite{U6}.

The full nuclear stopping is reached most possibly in central
collision. We use the effective centrality which is defined with
$\tilde{b} = b/b_{max}$ to replace the impact parameter b. Here
$b_{max}$ is the maximum for minimum bias events, corresponding to
the tip-tip and body-body UU collisions are about 13 $fm$ and 17
$fm$, respectively.

\begin{figure}[tbph]
\resizebox{!}{50mm}{\includegraphics{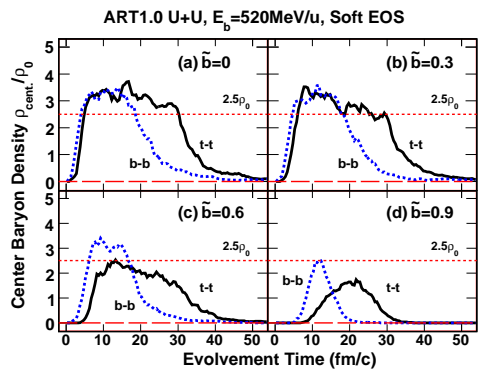}}
\caption{\label{Fig.2} Time evolution of central local baryon
density in UU tip-tip (solid line) and body-body (dashed line)
collisions with a beam kinetic energy of 0.52 GeV/u in four
different centralities.}
\end{figure}

Fig.2 shows the evolution of central baryon density in UU tip-tip
and body-body collisions  at $E_{\rm beam} = 0.52 GeV/u$ at
different centralities $\tilde{b}$ using the soft nuclear equation
of state with $K = 200$ MeV. Here, the central region is a cube with
$1fm^{3}$ and the lattice estimate is done for central
density~\cite{Q8}.

As expected, the maximum baryon density decreases evidently from
central to peripheral UU collisions for both tip-tip and body-body.
It is remarkable that the decline is faster and more obvious in
tip-tip than in body-body. The orientation of UU collision almost
has no effect during the early stage of the central collision
corresponding to $\tilde{b}\leq0.3$ when the kinetic energy is much
higher than the potential energy. A maximum baryon density of about
3.2 $\rho_{0}$ are reached at about 7 fm/$c$ in both tip-tip and
body-body collisions at $\tilde{b}=0$. The matter in the high
density region (i.e., with $\rho/\rho_{0}\geq2.5$ that may occur the
full nuclear stopping) lasts for about 13 fm/c and 25 fm/$c$ at
body-body and tip-tip collisions, respectively. This means that the
high density phase in the tip-tip collisions lasts about two times
longer than that in the body-body collisions. But this discrepancy
of high density phase lifetime fades away rapidly from $\tilde{b}=0$
to $\tilde{b}=0.3$, even the high density phase vanishes for tip-tip
collisions at about $\tilde{b}=0.6$. The higher compression and
longer passage time render the central tip-tip UU collisions as the
most probable candidates to form the QGP and to study nuclear EoS in
high density condition.

One should take note of that the tip-tip and body-body UU collisions
can reached a near maximum baryon density but exits a different
lifetime of high density matter for central events (See Fig2(a)). In
order to confirm what events are the interesting events with
extended lifetime, this discrepancy of tip-tip and body-body must be
displayed quantificationally. We define a new variable --- the
baryon aggradation strength as followed:
\beqar  
S = \int_{t^{'}}^{t^{''}}\rho dt, \eeqar Here, $t^{'}$ and $t^{''}$
are the threshold and end time of the central high density matter.
The centrality dependence of baryon aggradation strength is showed
in Fig.3. One can see clearly that the strength of tip-tip is higher
than body-body at $\tilde{b}\leq0.4$. The curve for tip-tip is flat
from $\tilde{b} = 0$ to $\tilde{b} = 0.2$ but has a sharp drop at
$\tilde{b} = 0.2$. Otherwise, the high density matter vanishes for
tip-tip but still exist for body-body at $\tilde{b}\geq0.7$.
Therefore, as a conservative estimate, the tip-tip events with
$\tilde{b}\leq0.2$, corresponding to $b\leq2.6fm$ and the value of
$S\geq70$, are the interesting events with extended high density
lifetime.

In Fig.3, we also indicate the result of using the forward (i.e.
$\theta<15^{0}$) neutron multiplicity $N_{n}$  to replace the
normalized impact parameter $\tilde{b}$ because the multiplicity of
forward neutron which can been measured easily in experiment at CSR
and have a linear relation with the impact parameter ~\cite{U7}. One
can see clearly that the multiplicity of forward neutron in tip-tip
is smaller than in body-body at $\tilde{b}\leq0.2$. Hence, we can
separate easily the tip-tip and body-body events only by the cut of
$N_{n}\leq30$ when most tip-tip central events will survive.
\begin{figure}[tbph]
\resizebox{!}{50mm}{\includegraphics{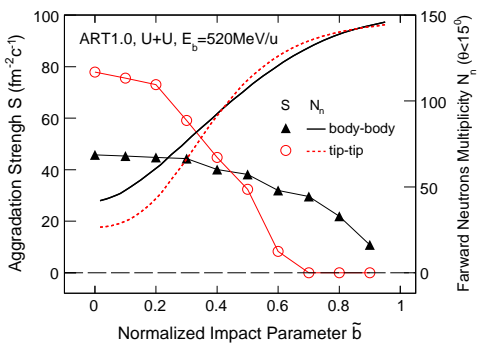}}
\caption{\label{Fig.3} Left: Baryon aggradation strength versus
normalized impact parameter for UU tip-tip (circular hollow dot) and
body-body (triangular solid dot) collisions. Right: The forward
neutron multiplicity versus normalized impact parameter for UU
tip-tip (dashed line) and body-body (real line)collisions at a beam
kinetic energy of $0.52 GeV/u$.}
\end{figure}

\section{\bf  The orientation dependence of central baryon density}
From Fig.1., one can see clearly that it depends strongly on five
parameters to confirm the initial geometries of UU collision. They
are the impact parameter $b$, the initial orientation of projectile
and target in reaction plane $\theta_{P}$ and $\theta_{T}$, the
inial orientation of projectile and target in transverse plane
$\phi_{P}$ and $\phi_{T}$, respectively.

In section II, we gained a simple estimate that the interesting
events with extended high density lifetime relate to $b\leq2.6 fm$.
Based on this point, we will discuss the orientation dependence of
central baryon density under $b\leq2.6 fm$ in this section.

Fig.4 (a) and (b) are the orientation of initial transverse plane
dependence on baryon aggradation strength at $b=0$ fm and $b=2.6$
fm, respectively. Here, we only need to take into account the
relative orientation in initial transverse plane because of the
symmetry of initial overlap region (interaction region) in shape.

\begin{figure}[tbph]
\resizebox{!}{50mm}{\includegraphics{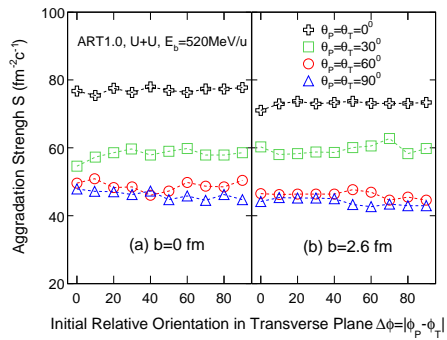}}
\caption{\label{Fig.4} The relative orientation of projectile and
target in the initial transverse plane dependence on baryon
aggradation strength at $b=0$ fm (a)and $b=2.6$ fm (b),
respectively. }
\end{figure}

From Fig.4, we can see that the baryon aggradation strength is
nearly invariable when the orientations of projectile and target
($\theta_{P}$, $\theta_{T}$) in the reaction plane are fixed but the
orientations of projectile and target in the transverse plane
($\phi_{P}$, $\phi_{T}$) are changed from $0^{0}$ to $90^{0}$ at
both $b=0$ and $b=2.6 fm$. Here, we only consider the extreme
condition that the orientations of projectile and target are equal,
i.e. $\theta_{P}=\theta_{T}$. There is an express drop is showed
from $0^{0}$ to $30^{0}$ for $\theta_{P}$ or $\theta_{T}$ when the
relative orientation in transverse plane $\Delta\phi$ is fixed. This
phenomenon illuminates both $\theta_{P}$ and $\theta_{T}$ should are
less than $30^{0}$ for the tip-tip like event.

\begin{figure}[tbph]
\resizebox{!}{60mm}{\includegraphics{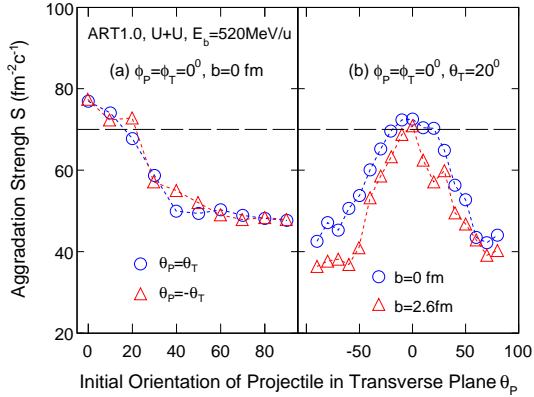}}
\caption{\label{Fig.5} The initial orientation of projectile and
target in reaction plane dependence on baryon aggradation strength
in UU collisions at $E_{b}=0.52 GeV/u$. Left window (a) with same
initial orientation $\theta_{P}=|\theta_{T}|$ at $b=0fm$. Right
window (b) with fixed $\theta_{T}=20^{0}$ at b=0 and $b=2.6fm$.}
\end{figure}

Due to inessential $\phi_{P}$ and $\phi_{T}$, fig.5(a) are the
initial orientation dependence on baryon aggradation strength at
$b=0$ fm with $\phi_{P}=\phi_{T}=0^{0}$. Hence, we only need to
consider two extreme geometries for full overlap in transverse
plane. Others are between them. The curve indicates the discrepancy
is tiny for $\theta_{P} =\pm\theta_{T}$ at $b=0fm$. It is remarkable
that a sharp drop at $\theta_{P} =|\theta_{T}|=20^{0}$ is seen. If
one requires $S\geq70$, the orientations in the initial transverse
plane should be less than $20^{0}$ for both projectile and target.
Fig.5 (b) shows $\theta_{P}$ dependence on baryon aggradation
strength with fixed $\theta_{T}=20^{0}$ at both $b=0fm$ and
$b=2.6fm$. It emphasizes again that $\theta_{P}$ is less than
$20^{0}$ if $S\geq70$.

In a short word, those events with extended high density phase in UU
collisions are the tip-tip like events which are with the initial
orientations $\theta_{P}\leq20^{0}$ and $\theta_{T}\leq20^{0}$ in
the transverse plane and $b\leq2.6fm$.

\section{\bf  Two measures for selecting out the tip-tip like events in experiment}
Due to the effect of initial orientation of projectile and target
for deformed nucleus, one key point to program UU collisions in
experiment is how to select out those collisions with the desired
geometry such as tip-tip like events.

In view of the discrepancy of forward neutron multiplicity $N_{n}$
dependence on impact parameter $b$ between tip-tip and body-body UU
collision showed in fig.3, a available try is associated
spontaneously. Fig.6(a) shows the forward neutron multiplicity
$N_{n}$ distribution with different selective impact parameters $b$
in random UU collisions. Here, About 6.8 millions events with
$b\leq6fm$ are created for our simulation. Obviously, it can
effectively reject those events with $b>4fm$ by selecting
$N_{n}<50$. Fig.6(b) shows impact parameter $b$ distribution with
different selective forward neutron multiplicity. If one selects the
cut with $N_{n}<40$, the percentage of those events with
$b\leq2.6fm$ will been enhanced. It is worth to mention that the
percentage can been farther enhanced with the cut $N_{n}<30$, but
large numbers of tip-tip like events with $b\leq2.6fm$ also will
been rejected. Thereby, $N_{n}$ is a good measure in experiment as a
trigger for fast rejecting those events with biggish impact
parameter $b$.

\begin{figure}[tbph]
\resizebox{!}{50mm}{\includegraphics{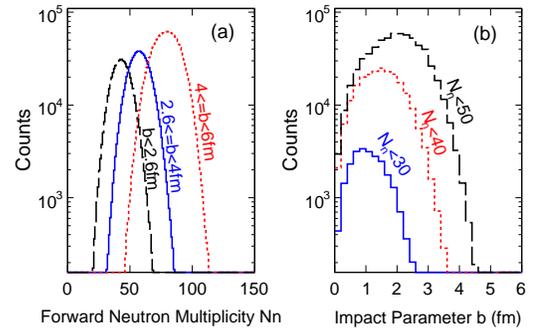}}
\caption{\label{Fig.6} Left window (a) is the forward neutron
multiplicity $N_{n}$ distribution with different selective impact
parameters $b$. Right window (b) is the impact parameter $b$
distribution with different selective forward neutron multiplicity
in random UU collisions at $E_{b}=0.52 GeV/u$.}
\end{figure}

As a incidental available result, fig.7 shows the initial
orientation ($\theta_{P}$ and $\theta_{T}$) distribution of
projectile and target in reaction plane with the cut $N_{n}<40$ and
$N_{n}<30$ for random UU collisions. One can see the most events is
tip-tip like events with both $\theta_{P}\leq20^{0}$ and
$\theta_{T}\leq20^{0}$. It is satisfying by the reducing $N_{n}$ to
gather more pure tip-tip like events.

\begin{figure}[tbph]
\resizebox{!}{45mm}{\includegraphics{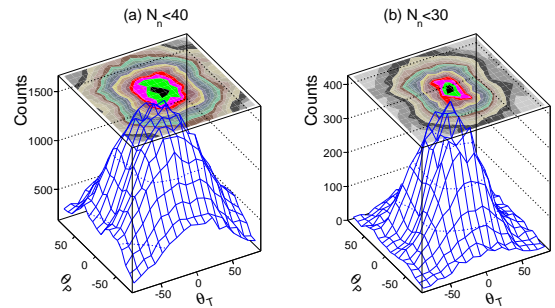}}
\caption{\label{Fig.7} The initial orientation ($\theta_{P}$ and
$\theta_{T}$) of projectile and target in reaction plane when cut
$N_{n}<40$ (a) and $N_{n}<30$ (b) in random UU collisions at
$E_{b}=0.52 GeV/u$.}
\end{figure}

In addition, it is well-know that an obvious long duration will load
possible thermal equilibrium of high density matter which has been
obtained in tip-tip central collisions. The degree of thermalization
can be measured by the ratio $R$ of transverse to longitudinal
momenta in low and intermediate energies heavy ion
collisions~\cite{U9}. Its expression is
\begin{equation}
R=\frac{2\sum_{i}|p_{it}|}{\pi\sum_{i}|p_{iz}|},
\end{equation}
Here, $i$ is the serial number of particles, the sum involves all
particles, $|p_{it} |$ and $|p_{iz} |$ are the total absolute value
of nucleon transverse and longitudinal momentum in the c.m.s,
respectively. It's a multi-particle observable on an event-by-event
basis, which for an isotropic distribution is unity. Then $R=1$ is a
necessary, although not is a sufficient, condition for thermal
equilibrium. It is also a measure of the stopping power~\cite{U10}.
\begin{figure}[tbph]
\centering
\includegraphics[width=0.5\textwidth]{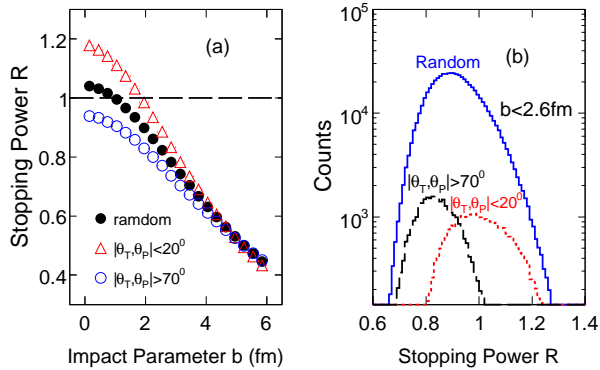}
\caption{\label{Fig.8} Left: the nuclear stopping power R as a
function of the impact parameter b in different orientation UU
collisions at $E_{b}=0.52 GeV/u$. Right: the nuclear stopping power
R distribution with $b<2.6fm$ in different orientation UU collisions
at $E_{b}=0.52 GeV/u$.}
\end{figure}

Fig.8(a) shows that the impact parameter b dependence on nuclear
stopping power R in different orientation UU collisions. Here, Only
charge particles are considered.  For three samples with different
orientation collisions, a nice linear relation can been seen clearly
between stopping power R and impact parameter b. For the most
central events, tip-tip like collisions ($\theta_{P,T}<20$) have a
maximal value for R among three samples, i.e. maximal most strong
stopping power. When $b$ increase, R decrease. Otherwise, that
stopping power R in tip-tip like collisions is large than in
body-body like events (($\theta_{P,T}>70$) at $b<4fm$. Even, R
reaches to $\sim$ 1.2 in tip-tip like events but only $\sim$ 0.92 in
body-body like events and $\sim$ 1.02 in random collisions at $b=0
fm$. Thus, we can gather the center events by selecting R, for
instance the cut with $R>1$ corresponding to $b\simeq1fm$ for random
and $b\simeq2fm$ for tip-tip like UU collisions can get rid of
body-body like events.

This remarkable discrepancy prompts us that the central tip-tip like
events maybe have a distinct distribution of stopping power R.
Fig.8(b) shows stopping power R distribution with $b<2.6fm$ for
three samples with different orientation UU collisions. It can be
see clearly that the body-body like events can been rejected
commendably from random events at $b<2.6fm$ by selecting $R>1$. If
one expects higher percentage of tip-tip like events in the
sub-sample, $R>1.1$ can been selected, but at the same time vast
tip-tip like events also will been thrown away.

\begin{figure}[tbph]
\centering
\includegraphics[width=0.5\textwidth]{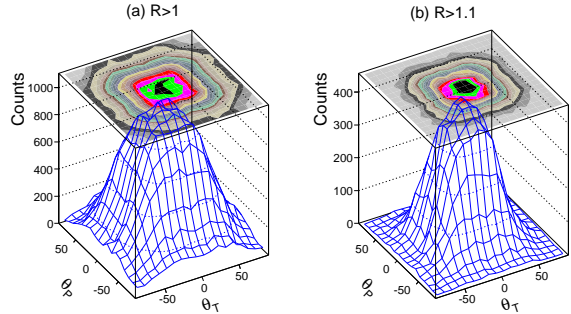}
\caption{\label{Fig.9} The initial orientation ($\theta_{P}$ and
$\theta_{T}$) of projectile and target in reaction plane when cut
$R>1$ (a) and $R>1.1$ (b) in random UU collisions at $E_{b}=0.52
GeV/u$.}
\end{figure}

Fig.9 shows the initial orientation ($\theta_{P}$ and $\theta_{T}$)
distribution of projectile and target in reaction plane when cut
$R>1$ (a) and $R>1.1$ (b) in random UU collisions. It can been seen
very clearly that most events have a small orientation, which events
are closer to tip-tip like events, by selecting the nuclear stopping
power $R>1$ and $R>1.1$. When $R>1.1$, the events with larger
orientation are rejected ulteriorly. In other words, it can enhance
the purity of tip-tip like events and minify the background when R
increases. Therefore, with the nuclear stopping power R, it is
possible to select out the tip-tip like collisions in experiment.
This will allow us to study the nuclear collisions at higher density
with a considerable long duration.

\section{\bf Summary}
Performing UU collisions are a good tool to search for QGP, the
phase boundary and the possible end-point of the phase boundary.
Those most interesting events (i.e. tip-tip like) is with small
initial orientations ($\leq20^{0}$) in reaction plane and small
impact parameter ($\leq2.6fm$). For the non-polarized UU collisions,
we have developed two measures allow us to select out those tip-tip
like events in experiment. The forward neutron multiplicity as a
trigger and nuclear stopping power is suited for fast on-line and
off-line physical analysis, respectively.

\vskip 1.cm \akgt \vskip 0.3cm
 This work is supported by NSFC of
china under the project 10775058, MOE of China under the contract
No. IRT0624, by the U.S. Department of Energy under Contract No.
DE-AC03-76SF00098, by CAS of china under KJCX2-SW-N18 and
CXTD-J2005-1.

\ed
\begin{thebibliography}{99}

\bibitem{Q1} R.Stock, {J Phys.},{\bf G30}: S633 (2004).
\bibitem{Q2} N.Xu, {Nucl.Phys.} {\bf A751}, 109 (2005).
\bibitem{U0} Lourencol, {Nucl. Phys.}, {\bf A698}: 13-22 (2002); H.Satz, {Nucl.
Phys.}, {\bf A715}: 3-19 (2003); R.Stock, {J. Phys.}, {\bf G30}:
S633-1423 (2004);
\bibitem{Q3} Workshop on 'Can we Discover the QCD Critical Point at RHIC', March
9-10, 2006; https://www.bnl.gov/riken/QCDRhic/
\bibitem{U1} E.K.Hyde, {Phys. Scr.} {\bf 10} 30-35 (1974) ;
\bibitem{U2} C.H¡§ohne, {Nucl. Phys.} {\bf A749},141c-149c (2005);
\bibitem{U3} P.Danielewicz, nucl-th/0512009;
\bibitem{U4} P.Danielewicz et al, {Science} {\bf 298},1592-1596 (2002);
\bibitem{U5} X.G.Li et al. {Nucl.Phys.Rev.} {\bf 243}, Vol.22 No.3(2005);
\bibitem{Q4} A.Bohr and B.Mottelson, {Nuclear Structure}, Vol. II (Benjamin, New York, 1975), p. 133.
\bibitem{S1} Memorandum written by P.Braun-Munzinger to BNL
management, of 9/18/1992.
\bibitem{U7} X.F Luo etc., {Phys.Rev.} {\bf C76}, 044902(2007);
\bibitem{Q5} B.A.Li, {Phys.Rev.} {\bf C61}, 021903 (2000).
\bibitem{Q6} E.V.Shuryak, {Phys.Rev.} {\bf C61}, 034905 (2000).
\bibitem{Q7} U.Heinz and A.Kuhlman, {Phys.Rev.Lett.} {\bf 94}, 132301 (2005).
\bibitem{U8} K.J Wu etc., {HEP \& NP}, {\bf 617} vol.31, No.7 (2007); K.J
Wu and F Liu, {HEP \& NP}, {\bf 1022} vol.31, No.11 (2007).
\bibitem{U9} H.Kruse etc., {Phys.Rev.} {\bf C31}, 1770 (1985).
\bibitem{U10} K.J Wu etc., {Chin.Phys.Lett.} {\bf 25}, 3204 (2008).
\bibitem{Q8} B.A.Li and C.M.Ko, {Phys.Rev.} {\bf C52}, 2037 (1995).
\bibitem{Q9} C.Y.Wong, Introduction to High Energy Heavy-Ion Collisions (World Scienti\_c, Singapore, 1994).
\bibitem{U6} P.F Zhuang, {Nucl.Phys.Rev.} {\bf 160}, Vol.16, No.6(2003).
\bibitem{Q10} S.A.Voloshin, {Nucl.Phys.} {\bf A715}, 379 (2003).
\bibitem{Q11} H.Sorge, {Phys.Lett.} {\bf B402}, 251 (1997).
\bibitem{Q12} J.Y.Ollitrault, {Phys.Rev.} {\bf D46}, 229 (1992).
\bibitem{Q13} N.Xu, Z.B.Xu, {Nucl.Phys.} {\bf A715}, 587 (2003).
\bibitem{Q14} X.Dong et al. {Phy.Lett.} {\bf B597}, 329 (2004).
\bibitem{Q15} H.H.Gutbrod et al., {Phys.Rev.} {\bf C42}, 640 (1990).
\bibitem{Q16} D.Brill et al., {Z.Phys.} {\bf A355}, 61 (1996).
\bibitem{Q17} D.Lambrecht et al., {Z.Phys.} {\bf A350}, 115 (1994).
\bibitem{Q18} N.Bastid et al., FOPI Collaboration, {Nucl.Phys.} {\bf A622}, 573 (1997).
\bibitem{Q19} J.Y.Ollitrault,Nucl-ex/9802005.

\end{thebibliography}
